\documentclass
[twocolumn,showpacs,preprintnumbers,superscriptaddress,amsmath,amssymb,prc]{revtex4}
\usepackage{graphicx}
\usepackage{dcolumn}
\usepackage{bm}

\begin{document}
\title{Different mechanism of two-proton emission from proton-rich nuclei $^{23}$Al and $^{22}$Mg}
\author{Y. G. Ma}
\thanks{ygma@sinap.ac.cn}
\author{D. Q. Fang}
\thanks{dqfang@sinap.ac.cn}
\author{X. Y. Sun}
\affiliation{Shanghai Institute of Applied Physics, Chinese Academy of Sciences, Shanghai 201800, China}
\author{P. Zhou}
\affiliation{Shanghai Institute of Applied Physics, Chinese Academy of Sciences, Shanghai 201800, China}
\author{Y. Togano}
\affiliation{Institute of Physical and Chemical Research (RIKEN), Wako, Saitama 351-0198, Japan}

\author{N. Aoi}
\author{H. Baba}
\affiliation{Institute of Physical and Chemical Research (RIKEN), Wako, Saitama 351-0198, Japan}
\author{X. Z. Cai}\author{X. G. Cao}\author{J. G. Chen}\author{Y. Fu}\author{W. Guo}
\affiliation{Shanghai Institute of Applied Physics, Chinese Academy of Sciences, Shanghai 201800, China}
\author{Y. Hara}
\author{T. Honda}
\affiliation{Department of Physics, Rikkyo University, Tokyo 171-8501, Japan}
\author{Z. G. Hu}
\affiliation{Institute of Modern Physics, Chinese Academy of Sciences, Lanzhou 730000, China}
\author{K. Ieki}
\affiliation{Department of Physics, Rikkyo University, Tokyo 171-8501, Japan}
\author{Y. Ishibashi}
\author{Y. Ito}
\affiliation{Institute of Physics, University of Tsukuba, Ibaraki 305-8571, Japan}
\author{N. Iwasa}
\affiliation{Department of Physics, Tohoku University, Miyagi 980-8578, Japan}
\author{S. Kanno}
\affiliation{Institute of Physical and Chemical Research (RIKEN), Wako, Saitama 351-0198, Japan}
\author{T. Kawabata}
\affiliation{Center for Nuclear Study (CNS), University of Tokyo, Saitama 351-0198, Japan}
\author{H. Kimura}
\affiliation{Department of Physics, University of Tokyo, Tokyo 113-0033, Japan}
\author{Y. Kondo}
\affiliation{Institute of Physical and Chemical Research (RIKEN), Wako, Saitama 351-0198, Japan}
\author{K. Kurita}
\affiliation{Department of Physics, Rikkyo University, Tokyo 171-8501, Japan}
\author{M. Kurokawa}
\affiliation{Institute of Physical and Chemical Research (RIKEN), Wako, Saitama 351-0198, Japan}
\author{T. Moriguchi}
\affiliation{Institute of Physics, University of Tsukuba, Ibaraki 305-8571, Japan}
\author{H. Murakami}
\affiliation{Institute of Physical and Chemical Research (RIKEN), Wako, Saitama 351-0198, Japan}
\author{H. Ooishi}
\affiliation{Institute of Physics, University of Tsukuba, Ibaraki 305-8571, Japan}
\author{K. Okada}
\affiliation{Department of Physics, Rikkyo University, Tokyo 171-8501, Japan}
\author{S. Ota}
\affiliation{Center for Nuclear Study (CNS), University of Tokyo, Saitama 351-0198, Japan}
\author{A. Ozawa}
\affiliation{Institute of Physics, University of Tsukuba, Ibaraki 305-8571, Japan}
\author{H. Sakurai}
\affiliation{Institute of Physical and Chemical Research (RIKEN), Wako, Saitama 351-0198, Japan}
\author{S. Shimoura}
\affiliation{Center for Nuclear Study (CNS), University of Tokyo, Saitama 351-0198, Japan}
\author{R. Shioda}
\affiliation{Department of Physics, Rikkyo University, Tokyo 171-8501, Japan}
\author{E. Takeshita}\author{S. Takeuchi}
\affiliation{Institute of Physical and Chemical Research (RIKEN), Wako, Saitama 351-0198, Japan}
\author{W. D. Tian}\author{H. W. Wang}
\affiliation{Shanghai Institute of Applied Physics, Chinese Academy of Sciences, Shanghai 201800, China}
\author{J. S. Wang}\author{M. Wang}
\affiliation{Institute of Modern Physics, Chinese Academy of Sciences, Lanzhou 730000, China}
\author{K. Yamada}
\affiliation{Institute of Physical and Chemical Research (RIKEN), Wako, Saitama 351-0198, Japan}
\author{Y. Yamada}
\affiliation{Department of Physics, Rikkyo University, Tokyo 171-8501, Japan}
\author{Y. Yasuda}
\affiliation{Institute of Physics, University of Tsukuba, Ibaraki 305-8571, Japan}
\author{K. Yoneda}
\affiliation{Institute of Physical and Chemical Research (RIKEN), Wako, Saitama 351-0198, Japan}
\author{G. Q. Zhang}
\affiliation{Shanghai Institute of Applied Physics, Chinese Academy of Sciences, Shanghai 201800, China}

\author{T. Motobayashi}
\affiliation{Institute of Physical and Chemical Research (RIKEN), Wako, Saitama 351-0198, Japan}

\date{\today}
\begin{abstract}
Two-proton relative momentum ($q_{pp}$) and opening angle ($\theta_{pp}$) distributions
from the three-body decay of two excited  proton-rich nuclei, namely
$^{23}$Al $\rightarrow$ p + p + $^{21}$Na and $^{22}$Mg $\rightarrow$ p + p + $^{20}$Ne,
have been measured with the projectile fragment separator (RIPS) at the RIKEN RI Beam Factory.
An evident peak at $q_{pp}\sim20$ MeV/c as well as a peak  in $\theta_{pp}$ around 30$^\circ$
are seen in the two-proton break-up channel from a highly-excited $^{22}$Mg. In contrast,
such  peaks  are absent for the $^{23}$Al case.
It is  concluded that the two-proton emission
mechanism of excited $^{22}$Mg is quite different from the $^{23}$Al case, with the former  having
a favorable diproton emission component at a highly excited state and the latter  dominated by the sequential decay process.
\end{abstract}
\pacs{23.50.+z, 25.60.-t, 25.70.-z, 27.30.+t}
\maketitle

{\it Introduction.---}
The decay of proton-rich nuclei, especially the two-proton (2$p$) radioactivity~\cite{gold60},
is an interesting process that may be observed in nuclei beyond or close to the proton
dripline \cite{pfutzner,blank08,Olsen}. Generally, there are two main ways for proton-rich nuclei to emit
two protons: (i) two-body sequential emission; (ii) three-body simultaneously
emission. But in the second way, there is an extreme case with the emission of two strongly correlated protons
(called 'diproton'). The diproton emission is basically two protons constrained by the pair correlation in a quasi-bound
$s$-singlet, i.e., $^1S_0$ configuration. Because of the Coulomb barrier, such a quasi-bound state can only exist for a short while and
then becomes separated after penetrating through the barrier.
Studying the two-proton correlation also
provides a good tool to understand the nucleon-nucleon pair-correlation (p-p correlation in particular)
inside a nucleus and  other related topics like the BCS-BEC crossover \cite{Shuck}. In addition, it is a good
way for investigating the astro-nuclear (2p,$\gamma$), and ($\gamma$,2p) processes which are closely related to the waiting
point nuclei \cite{astro1,astro2,astro3}.
Although some experimental investigations on the 2$p$ emitter have been done~\cite{kryger,gio,mukha,zerg,gomez,raci,lin,Wim,Ego},
the two-proton decay mechanism is still not well understood and further experimental and theoretical studies are required.

Kinematically complete decay channels of cold or low-excited nuclei can be reconstructed by advanced
detector arrays. For instance, the three-body decay channel of p + p + $^{A-2}_{Z-2}$Y from a proton-rich nucleus
$^A_Z$X can be identified by the Si-strip and other $\Delta E$ multi-detectors combination, which then allows
for the measurement of the opening angle, relative momentum and correlation function between two protons.
Since protons are not emitted chaotically in the two-proton decay, p-p coincidence measurements can, in principle,
deliver information of decay mode or nuclear structure, especially for proton-proton correlation of the parent nucleus \cite{Bertulani}.
As mentioned above, diproton emission is of interest. In this case, a strong correlation of p-p relative momentum around 20 MeV/$c$
will emerge together with a small opening angle between the two protons in the rest frame of the three decay products
as demonstrated in the experimental studies of $^{17,18}$Ne \cite{zerg,gomez,raci}.

Generally, the diproton emission process from the ground state is rare. If the lifetime is long enough, this is also called
two-proton radioactivity which was observed in a few nuclei~\cite{pfutzner,blank08}.
Two-proton radioactivity is predicted to occur for the even-$Z$ nuclei, for which, due to the pairing
force, one proton emission is energetically forbidden, whereas two-proton emission is allowed.
As this type of two-proton emission is essentially governed by the Coulomb and centrifugal barriers,
a sizable lifetime, which is compatible with the concept of radioactivity, is expected only for nuclei
with a reasonably high Coulomb barrier.
On the other hand, diproton emission itself is a more general phenomenon, especially for excited
states in proton-rich nuclei
since the decay is less suffered by the Coulomb barrier.

The proton-rich nucleus $^{23}$Al has also attracted a lot of attention in recent years since it may play a crucial
role in understanding the depletion of the NeNa cycle in ONe novae~\cite{iacob,gade,wies88}.
The measurement of its reaction cross section and fragment momentum distribution has shown that
the valence proton in $^{23}$Al is dominated by the $d$ wave but with an enlarged core~\cite{cai,fang}.
The spin and  parity of the $^{23}$Al ground state was found to be $J_\pi$= 5/2$^+$ \cite{ozawa,iacob}.
Also of great interest is $^{22}$Mg  because of its importance in determining the astrophysical reaction
rates for $^{21}$Na(p,$\gamma$)$^{22}$Mg and $^{18}$Ne($\alpha$,p)$^{21}$Na reactions in the explosive
stellar scenarios~\cite{wies99,sewe}.

In this Letter, we present an exclusive measurement to select the three-body decay
channels of $^{23}$Al and $^{22}$Mg, and investigate the relative
momentum and opening angle between the two protons. Based on the previous studies, a specific excitation energy window of 10.5 $< E^* <$ 15 MeV is used for $^{23}$Al and while 12.5 $ < E^* <$ 18 MeV for $^{22}$Mg, respectively. The window selections are based on (1) the data table of $^{23}$Al shows the existence of an excited state of 11.780 MeV
where  two-proton emission may exist~\cite{first};  (2) the transitions from the $^{22}$Mg ($T$=2) analog state (the excitation energy  is 14.044 MeV) to the ground state and/or first excited state
of $^{20}$Ne  was claimed but  they were unable to distinguish diproton emission or
sequential protons emission ~\cite{cable}.
Our results show a different two-proton
emission mechanism of $^{23}$Al and $^{22}$Mg as well as a clear diproton component from the decay
of $^{22}$Mg at high excitation energy, which demonstrates an interesting phenomenon.

{\it Experiments.---}
The experiment was performed using the RIPS beamline at the RI Beam Factory (RIBF) operated by RIKEN Nishina Center
and Center for Nuclear Study, University of Tokyo. The secondary $^{23}$Al
and $^{22}$Mg beams with incident energy of 57.4$A$ MeV and 53.5$A$ MeV, respectively, were generated
by projectile fragmentation of $135A$ MeV $^{28}$Si primary beam on $^{9}$Be production target and then
transported to a $^{12}$C reaction target. Around the reaction target, there was a $\gamma$ detector array
of 160 NaI(Tl) scintillator crystals named DALI2.  After DALI2 there were five layers of silicon detectors.
The first two layers of Si-strip (5mm width for one strip, 10 strips for one detector) detectors located around
50 cm downstream of the target were used to measure the emitting angle of the fragment and protons.
Three layers of  9 single-electrode Si were used as the $\Delta E$-$E$ detectors for the fragment. Each Si-strip layer
consists of 5$\times$5 matrix without detectors in the four corners. While each element Si layer consists
of 3$\times$3 matrix. Three layers of plastic hodoscopes located around 3 m downstream of the target were
used as $\Delta E$ and $E$ detectors for protons. Time-of-flight (TOF) of proton was measured by the first layer.
Most of the protons were stopped in the second layer.

The particle identification of $^{23}$Al and $^{22}$Mg before the reaction target was done by means of
$B\rho$-$\Delta E$-TOF method. After the reaction target, the heavy fragments were identified by five layers
of silicon detectors through the $\Delta E$-$E$ technique. Fragments with different charge and mass number are
well separated. Both the emission angle and energy loss can be obtained
for the fragments.  Total energy of    heavy fragments can be obtained by summing over the energy loss
of the five layers of silicon detector. Details about the experimental information can be found in Ref.~\cite{zhou}.
From this setup, a resolution better than 5 MeV/$c$ of the relative momentum
for protons at the typical energy of 65 MeV can be achieved.

\begin{figure}
\begin{center}
\includegraphics[width=9.2cm]{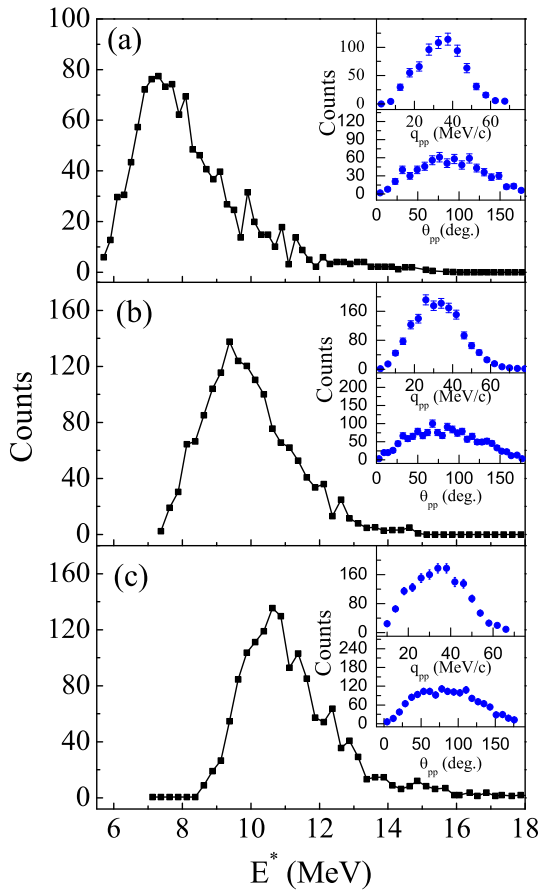}
\end{center}
\vspace{-0.8cm}
\caption{
(Color online) The excitation energy distributions constructed by the invariant mass of two-proton emission process
for $^{23}$Al (a), $^{22}$Mg (b) and $^{23}$Al $\rightarrow$ p + p + $^{20}$Ne (c).
The relative momentum and opening angle distributions of two protons are given in the inset of (a), (b) and (c), respectively.
}
\label{fig_exc}
\end{figure}

Clear particle identification were obtained for both the heavy fragments and protons.
The exclusive measurement for the break-up of the incident radioactive beam can be realized.
In our analysis, the (p + p + $^{A-2}_{Z-2}$Y) reaction channel can be picked and the
excitation energy of the incident nucleus $^{A}_{Z}$X can be reconstructed by the difference between the
invariant mass of three-body decay channel and mass of the mother nucleus in the ground state.
Fig.~\ref{fig_exc}(a) and Fig.~\ref{fig_exc}(b) show the excitation energy distribution obtained for the two proton
emission channel of $^{23}$Al and $^{22}$Mg, respectively.
Since the resolution for the reconstructed excitation energy  is estimated to be $\sim$1 MeV, it is difficult to
identify the specific excited states in $^{23}$Al and $^{22}$Mg.

\begin{figure}[t]
\begin{center}
\includegraphics[width=9.2cm]{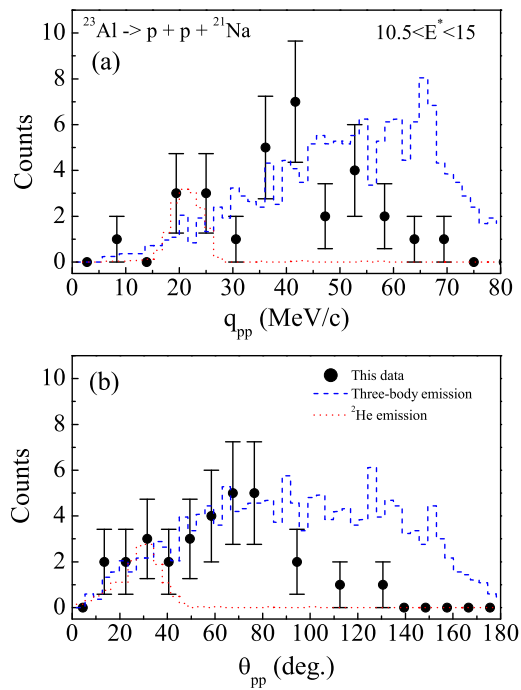}
\end{center}
\vspace{-2.8cm}
\caption{(Color online) Relative momentum distribution of two protons produced by the decay of
$^{23}$Al into two protons plus $^{21}$Na in the excitation energy window 10.5 $< E^* <$ 15 MeV (a);
Opening angle distribution between the two protons in the same excitation energy window (b).}
\label{fig_23Al-qpp}
\end{figure}

{\it Results and Discussion.---}
In the present study, we firstly examine the relative momentum spectrum ($q_{pp}$) and
opening angle ($\theta_{pp}$) of the two protons in the rest frame of three-body decay system
for odd-$Z$ nucleus $^{23}$Al and even-$Z$ nucleus $^{22}$Mg
without any cut in the excitation energy.
A broad $q_{pp}$ spectrum and structure-less $\theta_{pp}$ distribution are observed as
shown in the insets of Fig.~\ref{fig_exc}(a) and Fig.~\ref{fig_exc}(b).
These results indicate that the dominant mechanism of two proton emission from  $^{23}$Al and $^{22}$Mg
are sequential or simultaneous decay with weak correlation between the two protons.
Since the decay mode for different excited state or excitation energies could be  different, it will be interesting
to check $q_{pp}$ and $\theta_{pp}$ spectra in some excitation energy windows.
For diproton emission, a clear peak should appear at relative momentum around $\sim$ 20 MeV/$c$ as well as small opening angle.
Fig.~\ref{fig_23Al-qpp} shows the result of the above two distributions
for $^{23}$Al in excitation energy window 10.5 $< E^*<$ 15 MeV.  Evident peaks at $q_{pp}$ $\sim$ 20 MeV/$c$
(Fig.~\ref{fig_23Al-qpp}(a)) and smaller opening angle (Fig.~\ref{fig_23Al-qpp}(b)) are absent. Instead,
the $q_{pp}$ spectrum is broad and the $\theta_{pp}$ distribution is structure-less which are very similar to
the results of the whole excitation energy distribution.
Similar analysis has been checked in different E$^*$ windows other than 10.5 $< E^*<$ 15 MeV
and similar behaviors for  $q_{pp}$ and $\theta_{pp}$ are observed.


\begin{figure}[t]
\begin{center}
\includegraphics[width=9.2cm]{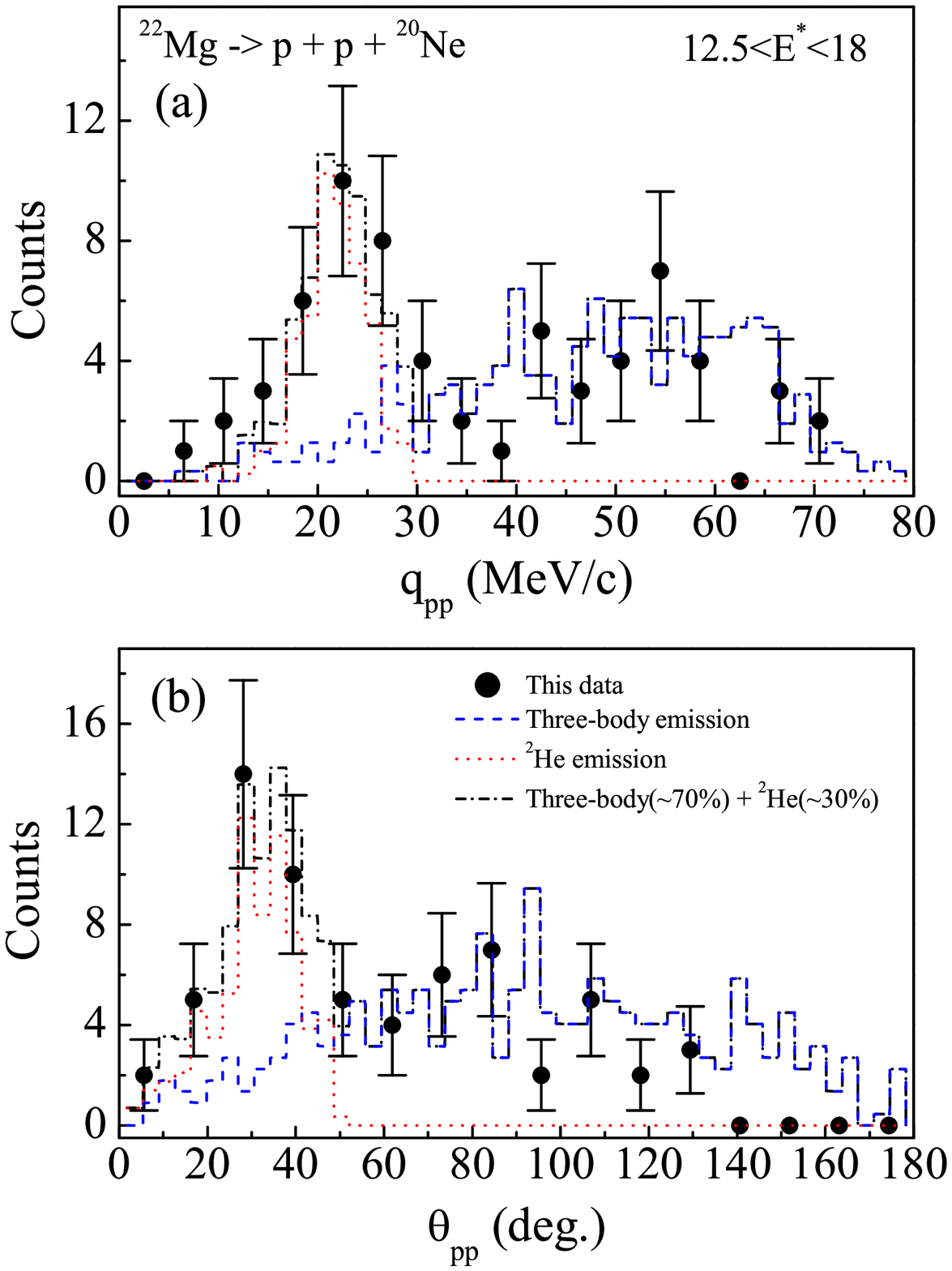}
\end{center}
\vspace{-2.8cm} \caption{(Color online) Same as Fig.~1 but for $^{22}$Mg in the excitation energy window
12.5 $< E^* <$ 18 MeV.} \label{fig_22Mg-qpp}
\end{figure}

Results have also been obtained for the even-$Z$ proton-rich nucleus, $^{22}$Mg. Fig.~\ref{fig_22Mg-qpp}
shows the relative momentum spectrum and opening angular distribution for the channel of
 p + p + $^{20}$Ne in the excitation energy window 12.5 $ < E^* <$ 18 MeV.
The peaks of the relative momentum distribution at 20 MeV/$c$ (Fig.~\ref{fig_22Mg-qpp}(a)) and
of the corresponding smaller opening angle (Fig.~\ref{fig_22Mg-qpp}(b)) are clearly observed.
These features are consistent with the diproton emission mechanism.
However, no significant enhancements for $q_{pp} \sim $ 20 MeV/c and small $\theta_{pp}$
are observed for other $E^*$ windows, which illustrates that the importance of the specific
window 12.5 $ < E^* <$ 18 MeV for diproton emission of $^{22}$Mg.

In order to quantitatively understand the $q_{pp}$ and $\theta_{pp}$
spectra, Monte Carlo simulations have been performed.
As shown in Fig.\ref{fig_exc}, the excitation energy spectrum is almost continuous,
it is difficult to distinguish the sequential decay  from the weak correlation simultaneous emission in our measurements.
Only two extreme cases are considered, i.e., diproton and weak correlation simultaneous three-body decay.
In the Monte Carlo simulation for $^{22}$Mg, the diproton decay spectrum was obtained by randomly sampling
the phase-space of the two-step process, $^{22}$Mg $\rightarrow$ $^2$He + $^{20}$Ne $\rightarrow$ p + p + $^{20}$Ne,
with the constraints of energy and momentum conservation and diproton being in the singlet-$S$ resonant
of two protons ($^2$He). The relative energy of the diproton was simulated according to Ref.~\cite{Ohn}.
The simultaneous three-body decay was simulated in the same way except that the
phase-space of the three-body p + p + $^{20}$Ne is sampled with only
the constraints of energy and momentum conservation. In Fig.~\ref{fig_23Al-qpp} and Fig.~\ref{fig_22Mg-qpp},
we show the diproton component by the dotted line and the three-body component by the dashed line.
As shown in Fig.~\ref{fig_23Al-qpp},  no trace for diproton emission is visible for $^{23}$Al as discussed before.
For $^{22}$Mg, on the other hand, the diproton emission peaks are well reproduced by the simulation.
The dash-dotted histograms in Fig.~\ref{fig_22Mg-qpp} represent the mixing of the two components.
The fraction of the diproton emission is about 30\%.
In similar previous experiments, around 70\% diproton emission contribution from highly excited $^{17}$Ne
was deduced~\cite{zerg} and around 30\% diproton emission contribution from the 6.15 MeV (1$^{-}$) state of $^{18}$Ne was observed~\cite{raci}.

Even though our excitation energy data is not precise enough to identify the exact excited state, the selected excitation
energy window 10.5 $< E^* <$ 15 MeV covers the 11.780 MeV excited state of $^{23}$Al ~\cite{first}.
Our observation illustrates that diproton emission is not visible in $^{23}$Al.
Since $^{23}$Al is an odd-$Z$ proton-rich nucleus, the diproton emission is
relatively difficult in comparison with the even-$Z$ proton-rich nucleus $^{22}$Mg.
In a previous $\beta$-delayed proton emission experiment for $^{22}$Al,
two-proton emission has been established but the decay mechanism is uncertain~\cite{cable}.
Our data confirm that there indeed exists diproton emission
(two-protons coupled to a $^1S_0$ configuration) by the observation of the peak at
$q_{pp}\sim20$ MeV/$c$ together with the small opening angles between the two protons
only in the excitation energy window 12.5 $< E^* <$ 18 MeV, which covers the 14.044 MeV state of
$^{22}$Mg with two-proton emissions.
On the whole, our present experiment definitely demonstrates 
that there exists a remarkable component of diproton emission process in the proton-rich nucleus
$^{22}$Mg.

\begin{figure}[t]
\begin{center}
\includegraphics[width=9.2cm]{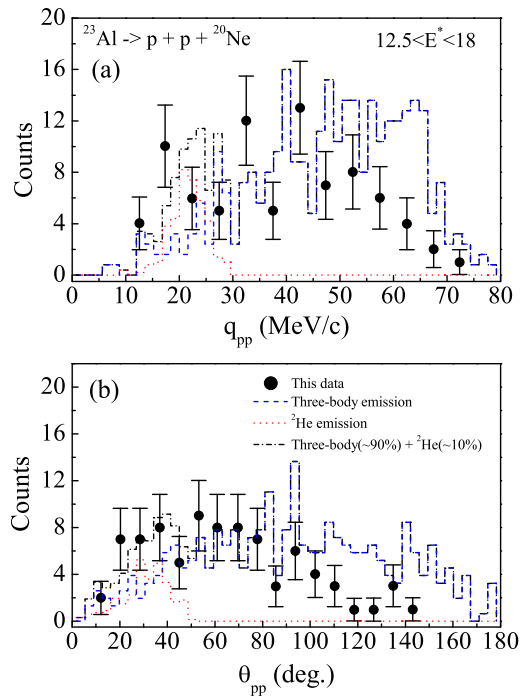}
\end{center}
\vspace{-2.8cm} \caption{(Color online) Same as Fig.~2 but for $^{23}$Al $\rightarrow$ p + p + $^{20}$Ne in the excitation energy window
12.5 $< E^* <$ 18 MeV.} \label{fig_23Al-22Mg-qpp}
\end{figure}

Considering excited $^{22}$Mg can also be produced by the single-proton removal from $^{23}$Al,
it provides us an alternative way to check the proton emission mechanism by
the decay process of $^{23}$Al $\rightarrow$ p + p + $^{20}$Ne,
where one proton was not detected in our experimental setup. In Fig.~\ref{fig_exc}(c), the excitation energy spectrum for
this process was shown together with the $q_{pp}$ and $\theta_{pp}$ distributions.
From the $q_{pp}$ and $\theta_{pp}$ spectra, a very small increase of statistics at $q=20$ MeV/c and small opening angle
can be seen.
To see more clearly, the relative momentum and opening angle distributions between two protons
in the excitation energy window 12.5 $< E^* <$ 18 MeV were shown in Fig.~\ref{fig_23Al-22Mg-qpp}.
A moderate enhancement appears at $q_{pp}$ $\sim$20 MeV/$c$ in Fig.~\ref{fig_23Al-22Mg-qpp}(a) and small angle in Fig.~\ref{fig_23Al-22Mg-qpp}(b),
which can be understood assuming the following two-step proton decay mechanism from $^{23}$Al.
First, one proton was emitted from $^{23}$Al and its corresponding residue nucleus is $^{22}$Mg.
Then other two protons are ejected from $^{22}$Mg and its corresponding residue nucleus is $^{20}$Ne.
Because of a remarkable 2$p$ correlation emission component in the second decay channel (Fig.~3),
a moderate 2$p$ enhancement could be eventually  observed in the process of $^{23}$Al $\rightarrow$ p + p + $^{20}$Ne.
The peak height of $q_{pp}$ in Fig.~\ref{fig_23Al-22Mg-qpp}(a) can be seen as a mixture of
Fig.2(a) and Fig.3(a), corresponding to events which have one proton from the first decay step
and another proton from the second decay step.
Actually a 10\% fraction of diproton emission can reproduce the data quite well as shown by
the dash-dotted histograms in Fig.~\ref{fig_23Al-22Mg-qpp}.

{\it Conclusions.---}
The measurements on two-proton relative momentum and opening
angle from the decay of the excited $^{23}$Al and $^{22}$Mg have been performed at the RIKEN RIBF.
In order to explore the internal proton-proton correlation information inside excited proton-rich nuclei,
decay channels of $^{23}$Al $\rightarrow$ p + p + $^{21}$Na and $^{22}$Mg $\rightarrow$ p + p + $^{20}$Ne
have been selected.  The results on the relative momentum and opening angle between the two protons are
presented. A broad $q_{pp}$ spectrum and structure-less $\theta_{pp}$ distribution are observed
for the whole excitation energy distribution which is reconstructed by the invariant mass method.
Peaks around $q_{pp} \sim $ 20 MeV/$c$ and $\theta_{pp} \sim 30^\circ$ are clearly observed  for the
even-$Z$ $^{22}$Mg at 12.5$<E^*<$18 MeV  covering the 14.044 MeV excited state with $T$=2,
which can be explained by a component of diproton emission.
For the odd-$Z$ proton-rich nucleus $^{23}$Al, the sequential decay is overwhelmingly dominant.
These results are confirmed by looking at the intermediate state of $^{22}$Mg in the process
of $^{23}$Al $\rightarrow$ p + p + $^{20}$Ne.

{\it Acknowledgements.---}
Authors are indebted to Che Ming Ko, Pawel Danielewicz and Carlos Bertulani for reading of the manuscript.
We are very grateful to all of the staffs at the RIKEN accelerator for providing beams during the experiment.
The Chinese collaborators greatly appreciate the hospitality from the RIKEN-RIBS laboratory.
This work is supported by the Major State Basic Research Development Program of China under contract
No. 2013CB834405, National Natural Science Foundation of China under contract
Nos 11421505, 11475244, 11035009 and 11175231.

\end{document}